\documentclass[preprint,aps,twocolumn,10pt]{revtex4}
\usepackage{amsmath,amssymb,amsthm,graphicx,latexsym}
\newcommand{\doublespacing}{\let\CS=\@currsize\renewcommand{\baselinesstrech}
{2.0}\tiny\CS}

\newcommand{\bd}{\begin{document}}
\newcommand{\ed}{\end{document}}
\newcommand{\bc}{\begin{center}}
\newcommand{\ec}{\end{center}}
\newcommand{\vs}{\vspace}

\begin{document}

\title{\Large \bf Scattering in a varying mass ${\cal{PT}}$ symmetric double heterojunction }

\vs{.3cm}

\author{{\bf Anjana Sinha}$^*$ \\
Department of Instrumentation Science, \\
Jadavpur University, Kolkata - 700 032, INDIA \\
and \\
{\bf R. Roychoudhury}$^ {\#}$ \\
Advanced Centre for Nonlinear and Complex Phenomena \\
1175 Survey Park,  Kolkata - 700075, INDIA }

\vspace{.5cm}

\begin{abstract}

\noindent We observe that the reflection and transmission coefficients of a particle within a double, $ {\cal{PT}}$ symmetric heterojunction with spatially varying mass, show interesting features, depending on the degree of non Hermiticity, although there is no spontaneous breakdown of ${\cal{PT}}$ symmetry. The potential profile in the intermediate layer is considered such that it has a non vanishing imaginary part near the heterojunctions. Exact analytical solutions for the wave function are obtained, and the reflection and transmission coefficients are plotted as a function of energy, for both left as well as right incidence. As expected, the spatial dependence on mass changes the nature of the scattering solutions within the heterojunctions, and the space-time (${\cal{PT}}$) symmetry is responsible for the left-right asymmetry in the reflection and transmission coefficients. However, the non vanishing imaginary component of the potential near the heterojunctions gives new and interesting results.

\vs{.5cm}

\noindent{\bf Key words :} Position-dependent-effective-mass; Rosen-Morse II;
${\cal{PT}}$-symmetry; Semiconductor Heterojunction; Scattering

\vs{.5cm}

\noindent{\bf PACS numbers :} 03.65.-w Quantum mechanics

\vs{.3cm}

\noindent $^*$ e-mail : anjana23@rediffmail.com; sinha.anjana@gmail.com \\
$^{\#}$ e-mail : rajdaju@rediffmail.com; rroychoudhury123@gmail.com \\
fax : +91 33 24146321; phone : +91 33 25753020

\end{abstract}

\maketitle

\newpage

\section{Introduction}

Position dependent effective mass (PDEM) formalism is extremely important
in describing the electronic and transport properties of quantum
wells and quantum dots, impurities in crystals, He-clusters,
quantum liquids, semiconductor heterostructures, etc.
[1-6]. In semiconductor heterostructures (e.g., say, Al$_{x}$Ga$_{1-x}$As), the spatial dependence on the mass
of the charge carrier (electron or hole) occurs due to its interaction with an
ensemble of particles within the device, as the particle propagates along the $z$
direction, because of the varying doping concentration or the mole fraction $x$
along the $z$-axis. On the other hand, it is established fact that non Hermitian
quantum systems with ${\cal{PT}}$ symmetry, open up a fascinating world, unknown
to conventional Hermitian systems [7-9].
${\cal{PT}}$ synthetic novel optical devices have been engineered
to exhibit several intriguing features [10-19]. These are structures with balanced gain and loss
such that the parity-time ($\cal{PT}$) symmetry of the entire system is preserved.
These materials depict altogether new behavior unknown to Hermitian optical systems ---
e.g., double refraction, power oscillations, unidirectional invisibility, left right asymmetry,
non reciprocal diffraction patterns, etc.

In this work we study a special form of
semiconductor device consisting of a thin layer of ${\cal{PT}}$
symmetric material sandwiched between two normal semiconductors,
such that the effective mass of the charge carrier (electron or hole) varies
with position within the heterojunctions, but is constant outside.
The mass $m(z)$ and the real part of the potential $V_R(z)$ are
taken to be continuous throughout the device. The form of $V(z)$
(where $V(z) = V_R (z) + V_I (z) $) is such that $V_I(z)$ does not
vanish near the heterojunctions. It is this non vanishing imaginary
component that gives rise to some interesting results. We obtain the
exact analytical solutions for the bound and scattering states of a particle
inside such a semiconductor device and also obtain the
reflection and transmission amplitudes, $R$ and $T$ respectively.
Our primary aim here is to look for any anomaly in $|R|$ and/or $|T|$,
with increase in the magnitude of the imaginary component $V_I(z)$.

\vs{.2cm}

The article is organized as follows : For the sake of
completeness, the position-dependent-mass Schr\"{o}dinger equation
is briefly introduced in Section 2, to show the procedure for
obtaining the exact analytical solutions. In Section 3, we study an
explicit non Hermitian ${\cal{PT}}$ symmetric double heterojunction,
the potential profile of which has a non vanishing imaginary part near the heterojunctions.
The potential and mass functions studied here are shown graphically, as function of $z$,
in Fig. 1. The exact analytical solutions for the scattering states of a particle in such a device are plotted in Fig. 2.
A complete understanding of any model requires knowledge of both bound as well as scattering states. Bearing this in mind, we plot the first three bound states in Fig. 3, for the same parameter values as in Figures 1 and 2.
The main stress in this work is on the behaviour of the transmission and reflection coefficients, with respect to the real and imaginary part of the potential, and the mass functions. To explore the phenomenon of left-right asymmetry, typical of non Hermitian quantum systems, a series of graphs showing the transmission coefficient $|T|^2$ and reflection coefficient $|R|^2$, for left and right incidence, are plotted in Figures 4 to 8. Section 4
is kept for Conclusions and Discussions.

\section{Theory}

\noindent Within the heterojuntions $a_1 < z < a_2$, where the particle mass varies with position, the
Hermitian kinetic energy term $T_{EM}$ is given by \cite{pdm4,harrison}
\begin{equation}\label{T-em}
    \begin{array}{lcl}
    T_{EM} &=& \displaystyle \frac{1}{4} \left( m^{\alpha} p
    m^{\beta} p m ^{\gamma} + m^{\gamma} p
    m^{\beta} p m ^{\alpha} \right) \\ \\
    &=& \displaystyle \frac{1}{2} p \left( \frac{1}{m}
    \right) p
    \end{array}
\end{equation}
where $ p = \displaystyle - i \hbar \frac{d}{dz} $ is the momentum
operator. The ambiguity parameters
$\alpha \ , \ \beta \ , \ \gamma $ obey the von Roos constraint
\cite{pdm4}
    \begin{equation}\label{abg}
        \alpha + \beta + \gamma = -1
    \end{equation}
For simplicity of calculations,
we shall work in units $\hbar = c = 1$, and use prime to denote
differentiation w.r.t. $z$.  Furthermore, for continuity
conditions at the abrupt interfaces, well behaved ground state
energy \cite{marrow, thomsen}, and the best fit to experimental results
\cite{proceed}, we shall restrict the ambiguity parameters to satisfy the
BenDaniel-Duke choice, viz., $ \alpha = \gamma = 0 \ , \ \beta =
-1 $. Thus, in the intermediate layer $a_1 <
z < a_2$, the Hamiltonian for the particle with PDEM assumes the form
\cite{plastino}
\begin{equation}\label{h-in}
        H = \displaystyle - \frac{1}{2m(z)} \frac{d^2}{dz^2} - \left(
        \frac{1}{2m(z)} \right) ^{\prime} \frac{d}{dz} + V_R(z) +
        i V_I (z)
\end{equation}
whereas, outside the well, $ z < a_1 $ and $ z > a_2 $, the
particle obeys the conventional Schr\"{o}dinger equation :
\begin{equation}\label{sch-out}
    \displaystyle \left\{ - \frac{1}{2m_{1,2}} \frac{d^2}{dz^2} +
    V_{01,02}
    \right\} \psi (z)  = E \psi (z)
\end{equation}
having plane wave solutions. In case we consider a wave incident
from left, the solutions in the two regions are
\begin{equation}\label{psi-out}
\begin{array}{lcl}
    \psi _L (z)  &=& \displaystyle e^{i k_1 z} + R e^{-ik_1 z} \ , \ - \infty <
    z < a_1 \\ \\
    \psi _R (z)  &=& T e^{i k_2 z} \ , \ \qquad \qquad a_2 <
    z < \infty \\
\end{array}
\end{equation}
where $R$ and $T$ denote the reflection and transmission
amplitudes, and
\begin{equation}\label{k}
    k_{1,2} = \displaystyle  \sqrt{ 2 m_{1,2} \left( E - V_{01,02} \right) }
\end{equation}
The important point worth noting here is that for PDEM systems,
the solutions $\psi (z)$ obey modified boundary conditions
\cite{benDaniel-Duke,boundary} --- the functions $\displaystyle \psi (z) $ and $\displaystyle
\frac{1}{m(z)} \frac{d \psi (z)}{dz} $ are continuous at
each heterojunction $a_1$ and $a_2$. These are used to calculate $R$ and $T$.

\vs{.5cm}

\noindent In case one uses the following transformations \cite{br-pr}
    \begin{equation}\label{psi-z}
         \psi _{in} (z) = \displaystyle \left\{ 2 m(z) \right\} ^{1/4} \phi
         (\rho) \  , \ \rho = \displaystyle \int \sqrt{2 m(z)} dz
    \end{equation}
then the Schr\"{o}dinger equation for PDEM in the region $a_1 < z < a_2$,
viz., eq. (\ref{h-in}), reduces to one for
constant mass viz.,
\begin{equation}\label{schro-const-m}
    \displaystyle -  \frac{d^2 \phi }{d \rho ^2} + \left\{
    \widetilde{V} (\rho) - E \right\} \phi  = 0
\end{equation}
with
\begin{equation}\label{v-tilde}
    \widetilde{V} (\rho) = \displaystyle V(z) + \frac{7}{32} \frac{m^{\prime
    \ 2}}{m^3} - \frac{m^{\prime \prime}}{8 m^2}
\end{equation}
Evidently, eq. (\ref{schro-const-m}) can be solved analytically
for some particular cases of $V(z)$ and $m(z)$ only.
In ref. \cite{as-PTpdm}, we had given the exact analytical solutions for one such case, which shows
the phenomenon of spontaneous ${\cal{PT}}$ symmetry breaking, and admits a spectral singularity. In this work, we shall study a second case,
which neither has any exceptional point in the bound state spectrum,
nor a spectral singularity in the continuous spectrum,
but nevertheless, shows some interesting results.
Additionally, contrary to our previous model, the most important contribution of the imaginary part of the
present potential is near the heterojunctions. Thus, this present study is distinctively different from the work done in ref. \cite{as-PTpdm}.

\section{Explicit model : ${\cal{PT}}$ symmetric potential well with position dependent effective mass}

\noindent We assume the real part of the intermediate layer to be a diffused
quantum well, similar to our earlier work \cite{as-PTpdm}. However, contrary to our earlier
study, the main contribution from the imaginary part of the potential is near the heterojunctions.
To be precise, we consider the following ansatz for the potential $V(z)$ and mass $m(z)$ :
\begin{equation}\label{pot-1}
        V(z) = \left\{
    \begin{array}{lcl}
        & & \displaystyle - \ \frac{\mu _1 }{ 1 + z^2  } + i
        \frac{\mu _2 z}{\sqrt{1+z^2} }
        \ \  , \ | z | < a_0   \\ \\
        & & \displaystyle - \frac{\mu _1 }{  1 + a_0^2
         } \ = \
        V_0 \qquad \ , \ | z | > a_0  \\
    \end{array}
    \right.
\end{equation}
\begin{equation}\label{mass-1}
        m(z) = \left\{
    \begin{array}{lcl}
        & & \displaystyle \frac{g ^2}{2 \left(1 + z^2 \right)}
        \ \qquad \  , \ | z | < a_0 \\ \\
        & & \displaystyle \frac{g ^2}{2 \left(1 + a_0^2 \right)} = m_0 \
        , \ | z | > a_0 \\
    \end{array}
        \right.
\end{equation}
where $\mu _1 , \ \mu _2 , \ g$ are some constant parameters.

\noindent Fig. 1 shows the mass dependence $m(z)$ and
the potential $V(z)$ in the entire semiconductor device,
as a function of $z$, for a suitable set of parameter values,
viz., $g = 1.5 , \ \mu_1 = 4, \ \mu_2 = .3, \ a_0 = 2.5 $.

{\begin{figure}[hp]
\begin{center}
\scalebox{0.6}{\includegraphics{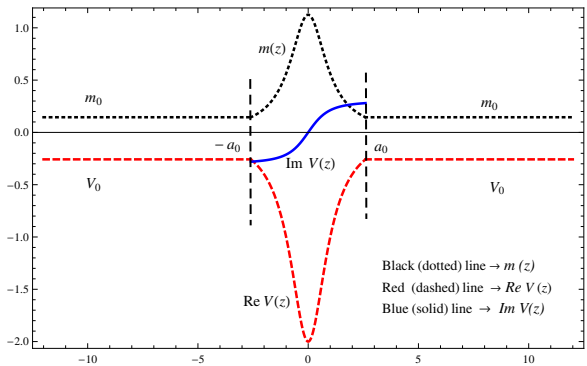}}
\label*{}\caption{\small {Colour online : Plot showing $m(z)$ and
$V(z)$ w.r.t. $z$ }}
\end{center}
\end{figure}}

\pagebreak

\noindent For the spatial mass dependence given by eq.
(\ref{mass-1}), eq. (\ref{psi-z}) transforms the coordinate $z$ to
\begin{equation}\label{rho}
    \rho = \beta \sinh ^{-1} z
\end{equation}
so that after some straightforward algebra $\widetilde{V} (\rho)$
in eq. (\ref{v-tilde}) reduces to
\begin{equation}\label{v-sech}
    \widetilde{V} (\rho) = \displaystyle \frac{1}{4 g ^2} -
    \left( \mu _1 - \frac{1}{g ^2}\right) {\rm{sech}} ^2   \frac{\rho}{g} + i
    \mu _2 g ^2  \ \tanh   \frac{\rho}{g}
\end{equation}
Thus equation (\ref{schro-const-m}) may be written as
\begin{equation}\label{schro-rho}
    \displaystyle \frac{d^2 \phi}{d \bar{\rho} ^2} + \left \{ \kappa ^2 +
    s \left( s + 1 \right) {\rm{sech}} ^2 \bar{\rho} - 2 i \lambda \
    \tanh \bar{\rho} \right\} \phi = 0
\end{equation}
\begin{equation}\label{k}
    {\rm{where}} \qquad \displaystyle \kappa ^2 =  E g ^2 - \frac{1}{4 }
    \ \ , \ \ \displaystyle \bar{\rho} = \frac{\rho}{g}
\end{equation}
and the new parameters $s$ and $\lambda$ are expressed in terms of the constants $\mu _1 ,
\mu _2 $ and $g$, as
\begin{equation}\label{lambda-s}
    \lambda = \displaystyle  \frac{1}{2} \mu _2 g^2 \qquad ,
    \qquad s = \displaystyle - \frac{1}{2} \pm g \sqrt{\mu _1}
\end{equation}
One can check from eq. (\ref{v-sech}) that the diffused potential well [eq. (\ref{pot-1})] within the intermediate layer $|z| < a_0$, with spatially varying mass $m(z)$,
reduces to the ${\cal{PT}}$ symmetric Rosen Morse II potential for constant mass \cite{levai-magyari,bikash-rr}. One may note that the standard ${\cal{PT}}$ symmetric Rosen Morse II potential has the following unique characteristics : \\
\ \ (i) absence of quasi-parity, \\
\ (ii) only real energy due to the absence of spontaneous breakdown of PT symmetry \\
(iii) switching of the bound state energies switch from
negative to positive values, with increase in the magnitude of the non Hermiticity parameter. \\
Our aim in this work is two fold --- \\
\ (i) to find the bound states of the system, check for exceptional points, and observe the effect of $\mu _2$, if any \\
(ii) to see the effect of $\mu _2$ on the behaviour of the reflection and transmission coefficients.

\vspace{.3cm}

To obtain the solution of (\ref{schro-rho}), we introduce a new variable
\begin{equation}\label{y}
    y = \displaystyle  \frac{1 - i \tanh  \bar{\rho}}{2}
\end{equation}
and write the solution as
\begin{equation}\label{phi-u}
    \phi = \displaystyle y^{\alpha /2} (1-y)^{\beta /2} \ \chi (y)
\end{equation}
After some straightforward algebra, equation (\ref{schro-rho})
reduces to the hypergeometric equation \cite{handbook}
\begin{equation}\label{y-u}
\begin{array}{lll}
    & & \displaystyle y(1-y) \frac{d^2 \chi}{dy^2} + \left\{
    \alpha  + 1 -  \left( \alpha + \beta + 2 \right) y \right\} \displaystyle \frac{d \chi}{dy} \\ \\
    & & - \displaystyle  \left\{ \displaystyle \left( \frac{\alpha + \beta + 1}{2} \right) ^2
    - \mu _1 g^2 \right\} \chi = 0 \\
\end{array}
\end{equation}
where $\alpha $ and $\beta $ are determined from the expressions
\begin{equation}\label{alpha-beta}
    \alpha ^2 + \kappa ^2 - 2 i \lambda = 0  \ , \
    \beta ^2 + \kappa ^2 + 2 i \lambda = 0
\end{equation}
Now, (\ref{y-u}) has complete solution \cite{handbook}
\begin{equation}\label{u-hypergeometric}
\begin{array}{lll}
    \chi &=& \displaystyle P \ _2F_1 \left( a,b,c; y
    \right) \\ \\
    &+& \displaystyle Q y^{1-c} \ _2F_1 \left( 1+a-c,1+b-c,2-c; y
    \right) \\
\end{array}
\end{equation}
where $P$ and $Q$ are constants, and the parameters $a$, $b$ and $c$
are as defined below :
\begin{equation}\label{ab}
\begin{array}{lcl}
    a &=& \displaystyle \frac{\alpha + \beta + 1}{2} + g \sqrt{\mu _1} \\ \\
    b &=& \displaystyle \frac{\alpha + \beta + 1}{2} - g \sqrt{\mu _1} \\ \\
    c &=& \alpha + 1 \\
\end{array}
\end{equation}
It is known from literature \cite{levai-magyari} that for bound states $Re (\alpha) > 0$ and $Re(\alpha) < 0$ are two mutually exclusive cases, described by solutions $ \phi ^+ (\bar{\rho}) $ and $ \phi ^- (\bar{\rho}) $, respectively. Additionally, regularity of the solution demands $Re(\beta) > 0 $. So, for bound states, we shall restrict ourselves to $ \phi ^+ (\bar{\rho}) $ only.

{\begin{figure}[hp]
\begin{center}
\scalebox{0.5}{\includegraphics{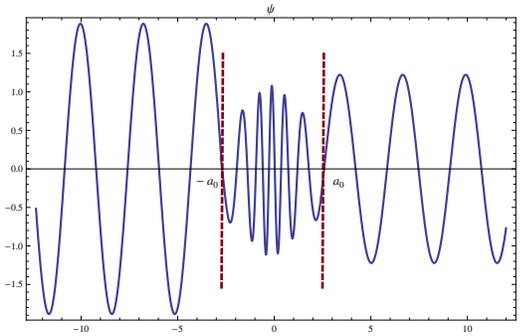}}
\label*{}\caption{\small {Colour online : A plot of Re $ \psi (z)$
vs $z$; Dashed (black) lines show the abrupt heterojunctions at $\pm a_0$ }}
\end{center}
\end{figure}}

After some straightforward algebra, the final solution
to the PDEM Schr\"{o}dinger equation within the
potential well $ | z | < a_0$, is obtained as
\begin{equation}\label{psi-in}
\begin{array}{lll}
    \psi _{in} (z) &=& \displaystyle \left(2 m \right) ^{1/4} y^{\alpha /2}
    (1-y)^{\beta /2} \ \left\{ P \ _2F_1 \left( a, b , c; y \right) \right. \\ \\
    &+& \displaystyle \left. Q y^{1-c} \ _2F_1 \left( 1+a-c,1+b-c,2-c; y
    \right) \right\} \\
\end{array}
\end{equation}
Outside the well ($ | z | >  a_0 $), the
scattering solutions are given by eq (\ref{psi-out}), with $ k_1 = k_2 = k \ (\rm{say})$, viz.,
$$ \begin{array}{lcl}
    \psi _L (z)  &=& \displaystyle e^{i k z} + R e^{-ik z} \ , \ - \infty <
    z < a_1 \\ \\
    \psi _R (z)  &=& T e^{i k z} \ , \ \qquad \qquad a_2 <
    z < \infty \\
\end{array} $$
while the bound states are given by
\begin{equation}\label{psi-bound}
    \begin{array}{lll}
    \psi ^{(b)} _{L} (z)  = \displaystyle A_1 e^{k_b z}  \ , \ - \infty <
    z < a_1 \\ \\
    \psi ^{(b)} _{R} (z)  = A_2 e^{- k_b z} \ , \  a_2 <
    z < \infty \\ \\
    \ \ \ \ \ \ k_b = 2 m_0 \sqrt{V_0 - E} \\
\end{array}
\end{equation}

To obtain the solution in the entire region, we need the expressions for the different coefficients --- $A_1$, $A_2$, $P$, $Q$,  $R$ and $T$. These are determined by applying the modified boundary conditions at each heterojunction $\pm a_0$, as mentioned earlier. Furthermore, we consider the various properties of the Hypergeometric functions $ _2F_1 \left( a, b , c; y \right) $ \cite{handbook}, and take the help of Mathematica. The scattering solution in the entire region is plotted in Fig. 2, for the same set of parameter values as in Fig. 1, viz., $g = 1.5 , \ \mu_1 = 4, \ \mu_2 = 0.3 , \ a_0 = 2.5 $, for $E=44$. Analogous to our previous studies on non Hermitian \cite{as-PTpdm} and Hermitian \cite{epl-anjana} models, we again observe that the dependence of position on the mass of the particle in the intermediate layer, changes the nature of the otherwise plane wave solution.

{\begin{figure}[hp]
\begin{center}
\scalebox{0.5}{\includegraphics{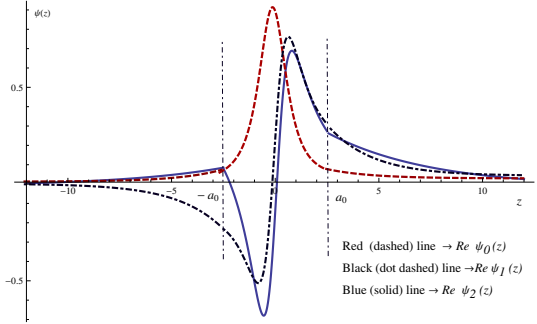}}
\label*{}\caption{\small {Colour online : A plot of Re $ \psi (z)$
vs $z$, for the first 3 bound states. Notice the kinks in the solutions at the abrupt heterojunctions at $\pm a_0$ }}
\end{center}
\end{figure}}

To have a full understanding of this model, we also calculate the bound state energy and eigenfunctions. For the same set of parameter values as in Figures 1 and 2, viz., $g = 1.5 , \ \mu_1 = 4, \ \mu_2 = 0.3 , \ a_0 = 2.5 $, we obtain the ground state at $E_0 = -8.82$, first excited state at $E_1 = -2.64$, and second excited state at $E_2 = -0.7$. The corresponding wave functions are plotted in Fig. 3. We observe a very interesting phenomenon --- the standard ${\cal{PT}}$ symmetric Rosen Morse II potential for constant mass and the varying mass diffused quantum well sandwiched between two heterojunctions, have some similar features : \\
 \ (i) The bound state energy is always real, hence there is no spontaneous breakdown of ${\cal{PT}}$ symmetry. \\
 (ii) In this particular case too, the bound state energy switches from negative to positive value, depending on the width of the potential well $2 a_0$, and the relative strengths of $\mu _1$ and $\mu _2$. \\
However, the bound state energies of Rosen Morse II and the present model are significantly different.
Additionally, for $\mu _1 = 4$, while the ground state energy switches from negative to positive value at $\mu _2 = 34.7222$ in case of the standard Rosen Morse II potential, for the model studied here, it occurs at $\mu _2 = 65.87$, for $g = 1.5 , \ \mu_1 = 4, \ a_0 = 2.5 $. Increasing the value of $a_0$ decreases this value of $\mu _2$ for energy switching.

As stated earlier, the main purpose of this work is to study the behaviour of reflection and transmission amplitudes. For this purpose, we plot a series of graphs showing $|R|^2$ and $|T|^2$, in different regimes. The system clearly depicts left-right asymmetry, typical of non Hermitian systems. While the transmission amplitude comes out to be the same for both left and right incidence $|T_L| = |T_R| = |T|$ (say), the case is quite different for the reflection amplitude $|R_R| \neq |R_L| $. Fig. 4 shows these values for $\mu _1 = 4$ and a low value of the non Hermiticity parameter, viz., $\mu _2 = 0.3$.
$|R_L|$ is normal ($|R_L| < 1$) when the particle enters from left --- the
absorptive side ($ V_I(z) < 0$), and anomalous ($|R_R| > 1$) when the particle enters from right
--- the emissive side ($ V_I(z) > 0$) \cite{as-PTpdm,zafar,cannata-annals}.

{\begin{figure}[hp]
\begin{center}
\scalebox{0.7}{\includegraphics{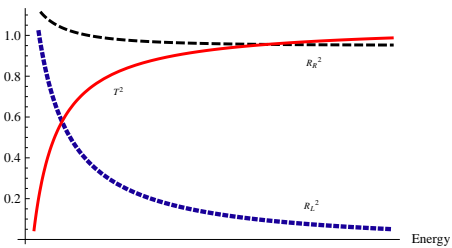}}
\label*{}\caption{\small {Colour online : A plot of $|T|^2$, $|R_R|^2$ and $|R_L|^2$ for different Energies, for $\mu _1 = 4, \ \mu_2 = 0.3, \ g = 1.5 , \ a_0 = 2.5$}}
\end{center}
\end{figure}}

As the non Hermiticity parameter $\mu_2$ increases, the behaviour of $|R_L|, \ |R_R|$ and $|T|$ changes abruptly.
For low values of $\mu _2$, for particle entering the device from left or right, $|T|$ increases with increasing energy, finally reaching unity --- total transmission. This observation is similar to that given in our earlier study \cite{as-PTpdm}. However, as $\mu _2$ increases, $|T|$ first decreases, reaches a minimum, and then increases to reach a saturation value. Once again, the pattern is identical for left and right incidence. This peculiar behaviour is shown in the 3D plot of Fig. 5. This abrupt change of behaviour occurs at a particular value of $\mu _2$, and the trend continues for all values of $\mu _2$ greater than this value.

{\begin{figure}[hp]
\begin{center}
\scalebox{0.6}{\includegraphics{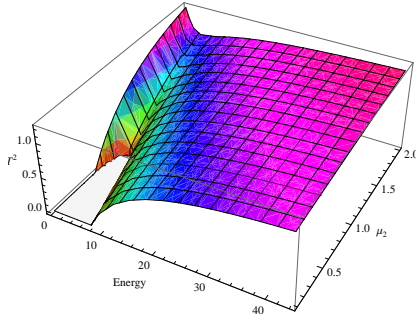}}
\label*{}\caption{\small {Colour online : A 3D plot of $|T|^2$, with respect to Energy and $\mu _2$, for $\mu _1 = 4, \ g = 1.5 , \ a_0 = 2.5$ }}
\end{center}
\end{figure}}

\noindent Similarly, if one draws the 3D plots for $|R_L|^2$ and $|R_R|^2$, with respect to energy and $\mu _2$, as shown in Fig. 6 and Fig. 7 respectively, once again there is an abrupt change in their behaviour at and beyond some critical value of $\mu _2$. The qualitative behaviour of $|T|^2$, $|R_L|^2$ and $|R_R|^2$ for large values of $\mu _2$, is shown in Fig. 8 (for $\mu _2 = 3$). This is in sharp contrast to their behaviour at low values of $\mu _2$, as shown in Fig. 4. However, the interesting point to note here is that the scattering coefficients remain finite everywhere, so the system does not exhibit spectral singularity. Thus, in spite of the absence of spectral singularity, the non Hermiticity parameter $\mu _2$ plays a crucial role in the behaviour of the scattering amplitudes, similar to its role in deciding the sign of bound state (negative or positive). These are the most important findings of the present study.

{\begin{figure}[hp]
\begin{center}
\scalebox{0.6}{\includegraphics{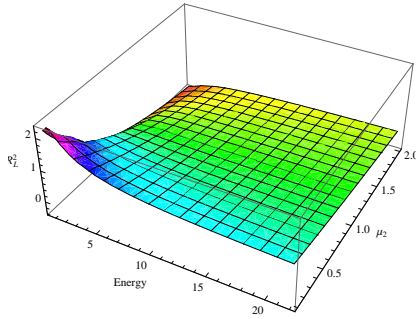}}
\label*{}\caption{\small {Colour online : A 3D plot of $|R_L|^2$, with respect to Energy and $\mu _2$, for $\mu _1 = 4, \ g = 1.5 , \ a_0 = 2.5$}}
\end{center}
\end{figure}}

Calculating the coefficients $|T|$, $|R_L|$ and $|R_R|$ in the limit $\mu _2 \rightarrow 0$, gives back the Hermitian results for these coefficients --- viz., $|R_L | = |R_R|$, and $ |T|^2 + |R|^2 \ = \ 1 $.


{\begin{figure}[hp]
\begin{center}
\scalebox{0.6}{\includegraphics{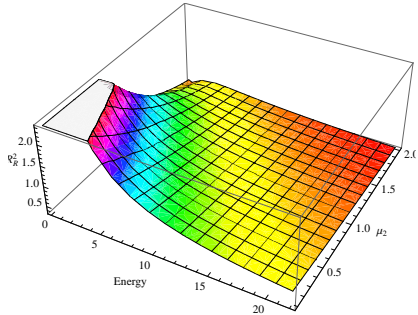}}
\label*{}\caption{\small {Colour online : A 3D plot of $|R_R|^2$, with respect to Energy and $\mu _2$, for $\mu _1 = 4, \ g = 1.5 , \ a_0 = 2.5$}}
\end{center}
\end{figure}}

{\begin{figure}[hp]
\begin{center}
\scalebox{0.5}{\includegraphics{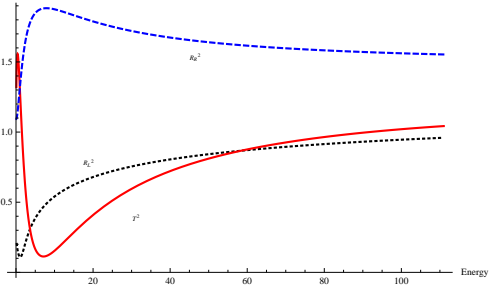}}
\label*{}\caption{\small {Colour online : A  plot of $|T|^2$, $|R_L|^2$ and $|R_R|^2$ with respect to Energy, for large $\mu _2$, viz, $\mu _2 = 3$}}
\end{center}
\end{figure}}

\section{Conclusions and Discussions}

To conclude, the special form of semiconductor device studied in this work displays some unique characteristics.
The particular model considered here does not undergo spontaneous breakdown of ${\cal{PT}}$ symmetry, nor it does not exhibit spectral singularity. Rather, the highlight of this PDEM device is the non vanishing imaginary part of the potential near the heterojunctions, within the intermediate layer.

The series of graphs plotted in the paper show the potential and mass functions (Fig. 1), the exact analytical scattering solutions in the entire device (Fig. 2), and also the first three bound state solutions (Fig. 3). The behaviour of the scattering amplitudes are shown in Figures 4 to 8. While the effect of the PDEM is to introduce a non linearity in the otherwise plane wave solutions (see Fig. 2), Fig. 4 shows the kinks at the heterojunctions. Numerical calculations show that the bound state energy switches from negative to positive value as $\mu _2$ increases.  Figures 4 to 8 give credence to the important role played by the non Hermiticity parameter $\mu _2$, in determining the scattering amplitudes. For low values of $\mu_2$, the nature of the reflection and transmission coefficients as shown in Fig. 4, is analogous to the observation in our previous study \cite{as-PTpdm}. However, as $\mu_2$ increases beyond a certain value, the qualitative picture of these coefficients changes abruptly. This is a new observation, hitherto unnoticed in earlier studies. At the same time this cannot be called a spectral singularity (ss), for the transmission and reflection coefficients viz., $|T|^2$, $|R_L|^2$ and $|R_R|^2$, blow up at a ss \cite{as-PTpdm,zafar,mostafazadeh-ss}. In this particular case, all the three coefficients remain finite. Additionally, their behaviour depends on whether the particle is entering the device from the left or from the right --- i.e., this non Hermitian system too possesses left-right asymmetry, despite the particle having PDEM in the region within the heterojunctions. For the particle entering the semiconductor device from the absorptive side ($V_I (z) \ < \ 1$), reflection is normal ($ |R_L| < 1 $), while for the particle entering the device from the emissive side ($V_I(z) > 0$) the reflection is anomalous ($|R_R| > 1 $). At the same time $ |T|^2 + |R|^2 \ \neq \ 1 $. In a fairly recent work it has been shown that for the ${\cal{PT}} $ symmetric Scarf II potential, in a particular regime, $ |T|^2 + |R_L| | R_R| \ = \ 1 $ \cite{zafar-2013}. However, it did not consider spatially varying mass, nor any abrupt heterojunction. In our present study of a ${\cal{PT}}$ symmetric heterojunction in the form of a diffused quantum well with PDEM, this conjecture is not valid.

With increase in artificial ${\cal{PT}}$ symmetric artificial optical structures, and semiconductor devices with position dependent mass heterojunctions, it is anticipated that this work provides some valuable insight into the transport properties of such a device, when a particle enters the material from one end and leaves from the other.

\section{Acknowledgement}

The authors thank B. Roy and B. Midya for some useful comments. One of the authors (AS) acknowledges financial assistance from the Department of Science and Technology, Govt. of India, through its grant SR/WOS-A/PS-11/2012.

\end{document}